\documentclass{aa}
\usepackage{graphics}
\usepackage{latexsym}

\def\EE#1{\times 10^{#1}}

\def\kms{\rm ~km~s^{-1}}

\def\ergs{\rm ~erg~s^{-1}}

\def\HI{{\rm H\,I}}

\def\lsim{\!\!\!\phantom{\le}\smash{\buildrel{}\over
  {\lower2.5dd\hbox{$\buildrel{\lower2dd\hbox{$\displaystyle<$}}\over
                               \sim$}}}\,\,}

\def\gsim{\!\!\!\phantom{\ge}\smash{\buildrel{}\over
  {\lower2.5dd\hbox{$\buildrel{\lower2dd\hbox{$\displaystyle>$}}\over
                               \sim$}}}\,\,}

\def\HI{H\,{\sc i}}

\def\degr{\hbox{$^\circ$}}
\def\arcmin{\hbox{$^\prime$}}
\def\arcsec{\hbox{$^{\prime\prime}$}}

\def\fs{\hbox{$.\!\!^{\rm s}$}}
\def\fdg{\hbox{$.\!\!^\circ$}}
\def\farcm{\hbox{$.\mkern-4mu^\prime$}}
\def\farcs{\hbox{$.\!\!^{\prime\prime}$}}

\def\la{\mathrel{\mathchoice {\vcenter{\offinterlineskip\halign{\hfil
$\displaystyle##$\hfil\cr<\cr\sim\cr}}}
{\vcenter{\offinterlineskip\halign{\hfil$\textstyle##$\hfil\cr
<\cr\sim\cr}}}
{\vcenter{\offinterlineskip\halign{\hfil$\scriptstyle##$\hfil\cr
<\cr\sim\cr}}}
{\vcenter{\offinterlineskip\halign{\hfil$\scriptscriptstyle##$\hfil\cr
<\cr\sim\cr}}}}}

\begin{document}

\thesaurus{11
              (11.03.1;  % Galaxies: clusters: general,
               11.03.4 MS2255.7+2039;  % Galaxies: clusters: individual,
               11.05.2;  % Galaxies: evolution,
               11.06.1;  % Galaxies: formation,
               11.12.2;  % Galaxies: luminosity function, mass function,
               12.03.3)} % Cosmology: observations

   \title{The Luminosity Function of MS2255.7+2039 at 
$z=0.288$\thanks{Based on observations made with the Nordic Optical Telescope,
              operated on the island of La Palma jointly by Denmark, Finland,
              Iceland, Norway, and Sweden, in the Spanish Observatorio del 
              Roque de los Muchachos of the Instituto de Astrofisica de 
              Canarias.}}

   \author{Magnus N\"aslund \and Claes
   Fransson \and Monica Huldtgren}
%   \offprints{M. N\"aslund}

   \institute{Stockholm Observatory,\\ SE-133 36 Saltsj\"obaden,\\ 
	Sweden}

   \mail{magnus@astro.su.se}
   \date{Received; accepted }

   \titlerunning{Luminosity Function of MS2255.7+2039}
   \authorrunning{M. N\"aslund et al.}
   \maketitle
%	\markboth{N\"aslund et al.: Luminosity Function of 
%MS2255.7+2039}{}
%
%
%1234567890123456789012345678901234567890123456789012345678901

\begin{abstract}
The luminosity function of MS2255.7+2039 at $z=0.288$ is determined down to a 
total magnitude of $R \sim 24$, corresponding to $M_{\mbox{\tiny{\rm R}}} 
\sim -17 \, ({\rm H}_{\mbox{\tiny{0}}} = 50 \ {\rm km \ s}^{-1} \ {\rm Mpc}^{-1})$. 
The data are corrected for incompleteness and crowding using detailed 
simulations. We find that the luminosity function is steeper than a standard 
Schechter function at faint magnitudes, and shows an excess of galaxies 
below $ M_{\mbox{\tiny{\rm R}}} \sim -19$. After corrections for light loss and crowding, the 
data can be described by a sum of two Schechter functions, one with 
$M_{\mbox{\tiny{\rm R}}}^{*} = -22.8$ and $\alpha = -1.0$, and one steeper 
with $M_{\mbox{\tiny{\rm R}}}^{*} = -18.9$ and $\alpha = -1.5$, representing 
the dwarf population. A straight-line fit to the faint part yields a slope similar to the Schechter $\alpha = -1.5$ of the dwarf population. The luminosity function of MS2255.7+2039 is compared 
to other clusters at lower redshifts, and does not show any significant 
difference. The redshift range for clusters in which increased number of dwarf 
galaxies have been found is therefore extended to higher redshifts. 
\end{abstract}

\section{Introduction} \label{sectintrod}

The faint end of the 
luminosity function (LF) is of great interest both in connection to the excess 
found in number counts of faint galaxies and for theories of cluster formation.
The latter is demonstrated especially in a series of papers by  Kaufmann et 
al. (1993), Kaufmann (1995a, b), Heyl et al. (1995) and Baugh et al. (1996), 
which show how the LF is related to the hierarchical clustering in e.g. the 
Cold Dark Matter (CDM) model. A steep faint end of the LF is unique not only to 
the CDM scenario, 
but is a generic prediction of hierarchical models of cluster formation
(White \& Frenk 1991). The difference of the LF in the field and in clusters of 
galaxies also  gives important information about environmental effects related 
to the galaxy formation process. In particular, ram pressure stripping and 
effects of interaction are likely to be more important for clusters than in 
the field (e.g., Moore et al. 1996). There is in this respect a clear 
connection to the Butcher-Oemler effect (Butcher \& Oemler 1984), seen for 
clusters, but not apparent in the field. The fraction of blue galaxies in 
clusters is larger at higher redshifts, which is often interpreted as an 
evolutionary effect. The evolution of the cluster LF is therefore of obvious 
interest.

The field LF has been investigated locally by several groups. Some of 
these report local LFs with flat faint-end slopes of $-1.1 \la \alpha \la 
-1.0$ (Loveday et al. 1992; Ellis et al. 1996), while others have found a
significantly steeper LF with $\alpha \la -1.5$ (Marzke et al. 1994; Lilly 
et al. 1995). 
At higher redshifts especially the CFRS survey (Lilly et al. 1995) and the 
Autofib survey (Ellis et al. 1996) give information about the 
LF in the field up to $z \sim 1$. Lilly et al. find that, 
while strong evolution is seen for the blue sample at $z \gsim 0.5$, 
the red LF is  changing little. At $z \sim 0.5$  the 
LF has brightened by $\sim 1$ magnitude in B. At higher 
redshifts the bright end stays constant, while the faint continues to 
increase, leading to a steepening of the faint end of the LF, from $\alpha = -1.1$ locally 
to $\alpha = -1.5$ at $z \simeq 0.5$, in broad agreement with the Autofib survey. A 
steepening of the LF can explain the excess counts found in deep surveys, as discussed 
by Gronwall \& Koo (1995).

While the field LF has been studied in fair detail, there have been only a 
few CCD investigations of the LF of clusters of galaxies until recently. 
Relatively nearby clusters, like Virgo, Fornax and Coma, have been studied 
by a number of authors (e.g., Bernstein et al. 1995, hereafter BNTUW; Lobo 
et al. 1997; Biviano et al. 1995; De Propris et al. 1995; Trentham 1998a). 
While some earlier investigations yield faint-end slopes of $\alpha \simeq 
-1.3$ (e.g., Ferguson \& Sandage 1988), more recent investigations, taking low 
surface brightness galaxies into account, point to steeper slopes, $\alpha 
\la -1.5$ (e.g., Bothun et al. 1991).

At higher redshifts the information about the cluster LF is scarce. The first 
detailed study was that by Driver et al. (1994b, henceforth DPDMD), who 
studied the R-band LF to $R = 24$ for the cluster Abell 963 at $z = 0.206$. 
While the high luminosity end can be well fitted with a Schechter function, 
they found an increase of faint galaxies between $-19 \lsim M_{\mbox{\tiny{\rm R}}} 
\lsim -16.5$, with a slope of $\alpha \simeq -1.8$. Further investigations at 
$0.1 \la z \la 0.2$ have strengthened the case for  
steep slopes, $\alpha \la -1.7$, at $-19 \la M_{\mbox{\tiny{\rm R}}}$ (Smith 
et al. 1997; Wilson et al. 1997, hereafter WSEC), although examples of flatter LFs also exist 
(Trentham 1998b). 

It is obvious that these results need confirmation for more clusters,  
especially at higher redshifts. Because of the strong evolution of the 
Butcher-Oemler effect, as well as from numerical simulations of cluster 
evolution (e.g., Kauffmann 1995a,b), one expects substantial evolution even
from $z = 0.2$ to $z = 0.4$. The detection of the faint end of the LF then becomes more difficult, both because the galaxies are fainter, and 
because of increasing contamination from the background. As a first step we 
here report observations of  the cluster MS2255.7+2039 at $z = 0.288$. We will 
in this paper use ${\rm H}_{\mbox{\tiny{0}}} = 50 \kms~Mpc^{-1}$ and $\Omega_{\mbox{\tiny{\rm M}}} = 1$.

\section{The data}

The coordinates of the center of MS2255.7+2039 = Zw 8795, hereafter MS2255, 
are $\alpha~=~22^{\rm h}~55^{\rm m}~40\fs6$, 
$\delta~=~20\degr~30\arcmin~04\farcs2$ (1950.0), and 
the redshift is $z = 0.288$ (Stocke et al. 1991). MS2255 was 
detected as an X-ray cluster in the Einstein Observatory Extended Medium 
Sensitive Survey (EMSS), with an X-ray luminosity of $2.0\EE{44} \ergs$ 
in the $0.3 - 3.5$ keV band (Gioia \& Luppino 1994), fairly 
typical for the EMSS selected sample. 

The galactic extinction can be estimated in several ways. Based on the \HI\ 
column density, the absorption in the direction of MS2255 ($l~=~90\fdg32, 
b~=~-34\fdg67$) is $A_{\mbox{\tiny{\rm B}}} = 0.18$ (Burstein \& Heiles 1982). With $A_{\mbox{\tiny{\rm R}}} = 
0.61~A_{\mbox{\tiny{\rm B}}}$ we get $A_{\mbox{\tiny{\rm R}}} = 0.11$. On the other hand, the estimated X-ray column 
density is $5.0\EE{20} \rm~ cm^{-2}$ (Gioia et al. 1990). With $N_{\mbox{\tiny{\rm H}}} = 
4.8\EE{21}~E_{\mbox{\tiny{\rm B-V}}}$ (Bohlin et al. 1978) and $A_{\mbox{\tiny{\rm R}}} = 2.4E_{\mbox{\tiny{\rm B-V}}}$ this 
gives $A_{\mbox{\tiny{\rm R}}} = 0.25$, considerably higher than the value deduced from the \HI\ 
column density. Finally, the recently presented COBE/DIRBE - IRAS/ISSA dust 
maps (Schlegel et al. 1998) yield $E_{\mbox{\tiny{\rm B-V}}} = 0.06$, which leads to $A_{\mbox{\tiny{\rm R}}} = 
0.14$. We will in this paper use $A_{\mbox{\tiny{\rm R}}} = 0.14$, but note that this may be in 
error by $\sim 0.1$ magnitude.

\subsection{Observations}

The data were obtained with the 2.56 m Nordic Optical Telescope and the 
Andalucia Faint Object Spectrograph (ALFOSC) in June 1997. The ALFOSC 
contained a thinned, back-side illuminated Ford-Loral 2K$^{2}$ chip 
that yielded a field of $6\farcm5 \times 6\farcm5$ and an image scale of 
0\farcs189/pixel. Every data set consists of the usual 
bias, dark, twilight-flatfield, and standard-star images, beside the science 
frames.

To determine the LF of the cluster it is crucial to correct for the 
contribution from the field. A nearby field at $\alpha = 22^{\rm h} 54^{\rm m} 
55\fs04, \delta = 21\degr 32\arcmin 19\farcs00 \ (1950.0)$ was chosen in order
to have similar galactic latitude and extinction properties as the cluster. 
Furthermore, the  
seeing conditions during the observations turned out to be similar. All these factors are crucial
 to ensure that the foreground and background will be as similar to the 
cluster field as possible. Systematic errors may otherwise easily enter into 
the subtraction of the background. 

The total exposure time was 5400 seconds for both the cluster and the 
background field, respectively, divided into exposures of 900 seconds. During the 
observations we rotated the instrument by 90\degr \ and/or made offsets 
by 10\arcsec \ between different frames. This was done in order to suppress 
the influence of bad regions on the chip and to make it possible to create a 
night-sky flatfield from the object frames (e.g., Tyson 1988). A pointing 
error of the telescope at the time of the observations may have 
caused the centre of the cluster and the image field centre to differ 
slightly.

\subsection{Reductions}

The bias level was determined from the overscan of each frame. These 
values 
were used to scale a master bias that was subtracted from the frames. We 
then used the shifted, bias-subtracted science frames to create a night-sky 
flatfield by removing objects with a 'smooth-and-clip' (N\"aslund 1995), 
followed by an averaging of the frames. We corrected the 
large-scale gradients of the twilight flatfield by the night-sky master flat, 
and used the corrected twilight flatfield to flat the science frames.

A narrow strip along the edges had to be excluded in the frames due to the 
structure of the thinned CCD. The flat-fielded frames were sky-subtracted, 
corrected for atmospheric extinction, aligned and finally combined (see 
N\"aslund 1995). 
The rotation and shifting of the frames decreased the effective area of 
the combined image, so that the area of the background and cluster images 
became $5\farcm6 \times 5\farcm5 = 31.0 \Box$\,\arcmin \ and $5\farcm4 
\times 5\farcm6 = 30.2 \Box$\,\arcmin, respectively 
(the final detection areas were reduced somewhat in order to avoid edge 
effects; see below). The images 
were calibrated using standard stars observed at intervals during the night. 
The seeing in the combined images was in both cases $0\farcs85$ (FWHM). 
The background 
field is shown in Fig.~{\ref{FigBkg}}, while Fig.~{\ref{FigCl}} displays the 
cluster.

\begin{figure}
%\resizebox{\hsize}{!}{\includegraphics{MSSbkg900cf.eps}}
\vspace{88 mm}
\caption{The $5\farcm6 \times 5\farcm5$ background field image in the
R band. The total exposure time is 5400 seconds.}
         \label{FigBkg}
\end{figure}

The individual frames of the cluster and background fields were checked 
for 'internal consistency'. We selected a few common objects in the magnitude 
range 17-21 and determined their brightness both in FOCAS (Jarvis \& Tyson 
1981; Valdes 1982, 1993) and DAOPHOT (Stetson 1987). All objects were 
consistent within a few hundredths of a magnitude or better. The weighted 
average of the standard stars observations 
yielded an statistical uncertainty of $\pm 0.01$ magnitude.

\section{Analysis of the fields}

\subsection{Object catalogue}

The reduced fields were analyzed using the FOCAS package. As a limit to our detections we used a value of $2.5 
\sigma$ of the sky noise and a detection area $A$ of 20 pixels, which 
corresponds to the seeing. We then utilized the ordinary FOCAS procedure, 
including sky 
correction and splitting of multiple objects. The splitting procedure was 
checked by simulations and worked satisfactory in most cases. A region of 50 
pixels along the edges was excluded when counting galaxies, which decreased 
the effective area of the background field to $27.65 \Box$\,\arcmin \ and of 
the cluster field to $27.02 \Box$\,\arcmin. This procedure resulted in a 
catalogue of objects with a number of parameters such as isophotal R 
magnitudes $(m_{\mbox{\tiny{R}}}^{\mbox{\tiny{\rm iph}}})$ and intensity 
weighted first-moment radii $(r)$. A plot of the detected galaxies is shown 
in (Fig.~{\ref{Figmiphir1}}).

\subsection{Simulations}

A substantial part of the analysis consists of simulations in order to 
determine completeness levels, fraction of noise detections, and magnitude 
corrections. When we approach the level of 
the sky noise we will obviously lose some galaxies in the noise, as well as 
pick up false detections. This problem is for a given total magnitude  most 
severe 
for galaxies with large scale lengths and therefore low  surface brightness. 
As discussed by, e.g., DPDMD and WSEC, the determination of this 
factor is far from trivial, and unfortunately, model-dependent assumptions 
about the galaxies are necessary. In one approach, discussed by DPDMD, one 
generates artificial galaxies with fixed parameters shifted to 
different redshifts, to describe the surface brightness distribution. An 
alternative method, used by WSEC, is to use real galaxies at brighter 
magnitudes, typical for the population in the field, as templates, and then 
rescale these to fainter magnitudes. This approach has the advantage of using 
realistic brightness profiles. It, however, implicitly assumes that the 
relative fractions of the different morphological types are the same for faint 
magnitudes as for bright, and also that the brightness profiles of a given 
type is independent of luminosity, except for a normalization factor. 
Neither of these assumptions are obvious. 

We used a generalized version of the DPDMD method in this investigation; the 
objects were not only characterized by their brightness, but also by their 
scale size. 
Simulated exponential disks were added to the background-field image, and 
these artificial galaxies were then detected with the same criteria as were 
used for the real data. After detection we could determine the magnitude 
correction depending on the objects position in the magnitude-scale length 
diagram (see Section~{\ref{sec:magcorr}}).  

\begin{figure*}
%\center{\resizebox{13cm}{!}{\includegraphics{MSS6_900f.eps}}}
\center{\vspace{13cm}}
\caption{R-band image of MS2255.7+2039. The field covers $5\farcm4 \times 
5\farcm6$ and the total exposure time is 5400 seconds. North is up and East
is to the left.}
         \label{FigCl}
\end{figure*}

\subsubsection{Noise simulations}

First, a number of Poisson-noise frames were created, with a noise level 
corresponding to that of the data. We then used FOCAS to detect spurious 
features with different combinations of the upper sigma limit and 
detection area. The parameters we settled for, $\sigma = 2.5$ and $A = 
20$ pixels ($\approx$ seeing), yielded six noise detections in an area of 
$27 \Box$\arcmin. This amounts to about 0.2\% of the actual detections in 
the 
background field.

\subsubsection{Completeness}

To estimate the completeness we simulated exponential disks of different 
magnitudes, scale lengths and inclinations. For each set of parameters, 51 
exponential disks were generated in empty regions of the background-field 
image. The reason for positioning the simulated galaxies in empty regions 
was to isolate the detection completeness due to surface brightness, and treat 
other effects, like overlapping (cf. Section~{\ref{sec:crowding}}), 
separately. We used FOCAS with the same detection criteria as for the real 
data. The parameters of the detected artificial objects could then be 
extracted from the resulting catalogue. As mentioned below, and discussed by 
other authors (DPDMD, Trentham 1997), the possibility of detection, as well 
as the fraction of the total light recovered, varies with surface brightness, 
which in turn depends on scale length and inclination for a given magnitude. 
At fainter magnitudes, galaxies with short scale length are, as expected, more 
easily detected than those with long scale lengths. The simulations indicate 
that we detect all dwarf galaxies, modelled as exponential disks ($0.5 \la 
r_{\mbox{\tiny{\rm d}}} \la 2$ kpc, where $r_{\mbox{\tiny{\rm d}}}$ is the 
disk scale length), down to $R \sim 25$. 

An important complication is the possible presence of low surface 
brightness galaxies (LSBGs). These come in different flavours, such as the 
sample of blue Low-Surface-Brightness Galaxies of McGaugh \& Bothun
(1994) and the Giant Low-Surface-Brightness Galaxies of Sprayberry et
al. (1995). The central surface brightness distribution of galaxies is
not well known, although  research during the last decade has shed
more light on this particular type of object (Impey \& Bothun
1997). If LSBGs are peaked around 
$\mu_{\mbox{\tiny{\rm B}}}^{\mbox{\tiny{0}}} \simeq 23.2 \ 
\rm{mag}/\Box$\,\arcsec (corresponding to $\mu_{\mbox{\tiny{\rm
R}}}^{\mbox{\tiny{0}}} \simeq 22.0 \ \rm{mag}/\Box$\,\arcsec), as those
observed by Sprayberry et al. and McGaugh \& Bothun, we would detect
LSBGs with small and intermediate scale lengths 
($r_{\mbox{\tiny{\rm d}}} \sim 3$ kpc), given the colours of the Sprayberry 
et al. sample. Even objects with lower surface brightness, like UGC 9024 
($\mu_{\mbox{\tiny{\rm B}}}^{\mbox{\tiny{0}}} = 24.5, 
r_{\mbox{\tiny{\rm d}}} = 5.6/h$ kpc), should be possible to detect at 
$z = 0.288$, although such galaxies are close to our detection limit. 
Galaxies with lower surface brightness will accordingly escape 
detection. The most extreme cases known (e.g. Malin 1) are far beyond 
detection, but galaxies of this type are likely to be an order of 
magnitude fewer than objects of higher surface brightness (Davies et al. 1994).
Furthermore, a study of Hubble Deep Field data by Driver (1999) shows that 
luminous low surface brightness galaxies are rare compared to their high 
surface brightness counterparts. One should, in any case, bear in mind that 
LSBGs below the surface-brightness detection limits may influence the 
faint-end slope by an unknown amount.

\subsubsection{Magnitude correction}
\label{sec:magcorr}

From the simulations above, used to determine the completeness, 
we also estimated the total magnitude for each parameter set. During the 
analysis we developed 
a technique for magnitude corrections (N\"aslund 1998), which turned out to be 
similar to that of Trentham (1997). A large number of simulated galaxies, 
with different values of scale length ($r_{\mbox{\tiny{\rm d}}}$), axial 
ratio ($b/a$) and total magnitude ($m_{\mbox{\tiny{R}}}^{\mbox{\tiny{\rm 
tot}}}$), were generated in the background-field image. These objects were 
detected with the same setup as for the real data. We could at this 
stage calculate the magnitude correction as a function of isophotal magnitude 
($m_{\mbox{\tiny{R}}}^{\mbox{\tiny{\rm iph}}}$) and intensity weighted 
first-moment radius ($r$). We, finally, corrected the magnitudes of the detected 
galaxies from their position in the 
$(m_{\mbox{\tiny{R}}}^{\mbox{\tiny{\rm iph}}},r)$ plane (Fig.~{\ref{Figmiphir1}}). 

This method will obviously not be fully correct for elliptical galaxies that 
are better described by de Vaucouleur profiles than by exponential disks. 
Most of these are, however, of comparatively bright magnitudes, where the 
correction is small. If we assume that the faintest cluster members, for 
which the corrections are largest, have exponential profiles the application 
of the method is justified. This is plausible if the faintest galaxies are 
late-type spirals or dwarf spheroidals and/or dwarf irregulars. However, some 
of the galaxies close to our magnitude limit ($R \sim 24$) may be {\em 
luminous} dwarf ellipticals ($-16 < M_{\mbox{\tiny{B}}}$) that are better 
fitted by de Vaucouleurs profiles (Ferguson \& Binggeli 1994). We performed a 
few simulations of de Vaucouleurs profiles with short scale lengths to mimic 
luminous dwarf ellipticals, and compared them to exponential disks of the same 
brightness. We find that the total magnitudes for the $r^{1/4}$ profile 
galaxies are underestimated by 0.15 magnitudes, similar to the findings of 
Trentham (1997).  

The magnitude correction can be applied in two ways. One can either correct 
for 'light loss' without any assumptions about the cluster population, or one 
may use {\em a priori} information to constrain the distribution of points in 
the ($m_{\mbox{\tiny{R}}}^{\mbox{\tiny{\rm iph}}},r)$ plane. If the faintest 
cluster members that we detect are exclusively dwarfs, this has two 
implications. Firstly, for objects with $r_{\mbox{\tiny{\rm d}}} \la 2$ kpc 
the catalogue should be more than 50\% complete for 
$m_{\mbox{\tiny{R}}}^{\mbox{\tiny{\rm iph}}} < 26$. Secondly, the magnitude 
correction would in this case be fairly small, and the ambiguity at the 
faintest magnitudes is reduced, compared to a more complex population. This is 
because faint galaxies are found in a comparatively small region of the 
$(m_{\mbox{\tiny{R}}}^{\mbox{\tiny{\rm iph}}},r)$ plane, and the reason is 
simply that faint, intrinsically large galaxies (i.e. LSBGs), will have 
their apparent scale lengths substantially reduced, and thereby be closer to 
the true dwarf region of the $(m_{\mbox{\tiny{R}}}^{\mbox{\tiny{\rm iph}}},r)$ 
plane. As a result, the uncertainties in the magnitude correction 
increase at the faint end. If one for good reasons could justify the 
exclusion of non-dwarfs from this region of the 
$(m_{\mbox{\tiny{R}}}^{\mbox{\tiny{\rm iph}}},r)$ plane, the corresponding 
magnitude correction would be more accurate. In addition, if all faint 
galaxies are dwarfs and have $r^{1/4}$ profiles, the luminosity after 
correction would be systematically underestimated (see above). On the other 
hand, if they {\em all} are luminous dwarf ellipticals the 
magnitude correction would not be increasing with magnitude as steeply as the 
one applied here, and the LF would therefore be less steep at faint 
magnitudes. However, there is no strong motivation for such an exclusion of 
intrinsically larger objects, and for the data presented here we used the 
first, more general, approach. The possible presence of larger, faint objects 
also calls for the use of exponential profiles.

The magnitude-correction procedure was tested by generating a number of 
exponential disks distributed in magnitude according to a power law of 
slope 0.4. The 
galaxies were positioned along a grid in order to avoid crowding effects in 
this particular test. The simulations showed that the procedure managed to 
correct for the light loss well down to $R \simeq 24$, 
close 
to the actual completeness limit (for $r_{\mbox{\tiny{\rm d}}} \la 6$ kpc) 
of the data (N\"aslund 1998).

\begin{figure}
%\rotatebox{0}{\resizebox{\hsize}{!}{\includegraphics{miphir1.eps}}}
\vspace{88mm}
	\caption[]{The detected galaxies in the $(m_{\mbox{\tiny{R}}}^{\mbox{\tiny{\rm iph}}},r)$ plane of the cluster image. $m_{\mbox{\tiny{R}}}^{\mbox{\tiny{\rm iph}}}$ is the isophotal R-band magnitude and $r$ the intensity weighted first-moment radius.}
         \label{Figmiphir1}
\end{figure}

\subsubsection{Crowding}
\label{sec:crowding}

Overlapping objects is another important factor, especially for faint
objects. 
One way 
to see this is by noting that fainter objects have smaller effective 
field areas at their disposal. Stars and galaxies that are brighter in general 
occupy a larger apparent area in the image and fainter objects are shielded 
by them. 

We have tested different methods for this correction. In the first 
approach  we added simulated compact objects of different magnitudes to the 
cluster and background images. To a first approximation, the fraction
of recovered objects gives the detection probability as a function
of magnitude. However, non-linear effects, connected with obscuration
between especially the faint galaxies themselves, are likely to be
important, and full simulations of the cluster and background would be
needed to address these aspects. 

Instead of this method we adopted a more conservative correction procedure 
for the obscuration of faint objects. The cumulative area covered by objects 
up to a certain magnitude was calculated in half-magnitude intervals for the 
background and cluster field, respectively. It was then found that there is 
a clear difference between the area covered in the cluster image and the 
background~-~field image when it comes to {\em bright} objects. However, for 
fainter objects there is no substantial difference; the covered area increases 
similarly in both fields. In practice, this means that for objects brighter 
than $R \sim 22$ the obscured area is 8\% in the cluster image, while the 
corresponding number for the background image is only 3.4\%. We accordingly 
used these numbers to correct for the obscuration for objects fainter than 
$R = 22$. We do therefore not include any magnitude dependence of the 
correction 
factor below $R = 22$, as Smith et al. (1997) do, which implies that apparently small 
objects are not an important source of obscuration in our fields. Furthermore, these effects 
do not add linearly with brightness for faint objects.
In the case that faint objects contribute somewhat themselves to the 
obscuration, we would have underestimated the faint~-~end slope of the 
LF slightly. We hope to perform a more thorough study of crowding effects 
elsewhere.

\subsection{Foreground and background corrections}

\subsubsection{Stars}

MS2255 is at comparatively low galactic latitude ($b = -34\fdg67$), 
and a substantial contamination by stars is expected. Because our comparison
field is close to the cluster, this contamination is to a large extent 
reduced when we subtract the background counts from the cluster counts. 
With FOCAS, we can nevertheless separate stars from galaxies, based on the 
brightness profile, for $R \la 20$. This eliminates the statistical errors 
in the cluster counts from this source. The saturated stars were removed 
interactively from the galaxy list, while remaining stars brighter than 
$R = 20$ were detected and classified by FOCAS. We found in this way 25 
stars in the cluster field, and 17 in the background field. This can be 
compared to the number expected from the galactic model by Bahcall \& 
Soneira (1980) and Bahcall (1995), which for a field of $27 \ \Box\arcmin$ 
gives 30 stars brighter than $R = 20$. Within the statistical errors we 
consider the number of stars 
in the cluster field to be consistent with that expected from the model. The 
lower number for the background field is simply explained by the fact that 
the field was selected in a region void of bright objects. 

Besides bright stars, we also excluded small spurious objects
around bright stars or galaxies, which can be a result of false
detections by FOCAS in the cluster and background field (see Trentham
1997 for a discussion). This effect does not affect the faint-end slope
significantly.

\subsubsection{Background counts}

Guided by the completeness simulations, we decided to set the isophote 
limit for inclusion in the catalogue at $m_{\mbox{\tiny{R}}}^{\mbox{\tiny{\rm iph}}} 
= 26$, approximately corresponding to the magnitude limit after correction given above. 
According to the simulations we detect all pointlike sources at this isophotal 
magnitude. 

Inspection of the ($m_{\mbox{\tiny{R}}}^{\mbox{\tiny{\rm iph}}},r$) plane shows that some of 
the detected objects fall below the simulated point sources, i.e. they have 
scale lengths smaller than the seeing. The apparent small scale length may 
be a result of the small number of counts (ADUs) for the faint objects. There 
is then a substantial probability that even for a point-like object the scale 
length will be smaller than the width of the PSF. 
Moreover, the simulations showed that disks with large scale lengths that are 
below the completeness limit could be detected as such compact objects, but 
almost all of them have $m_{\mbox{\tiny{R}}}^{\mbox{\tiny{\rm iph}}}$ beyond our catalogue limit 
anyway, and hence cause no problem. The ambiguity in the interpretation of the 
nature of these objects made us test whether they influenced the shape of the 
cluster LF. We generated one list that included these objects, and one in which 
they were removed for both the cluster image and the background-field image. 
The resulting faint-end slope is in the latter case somewhat flatter compared 
to the slope including these objects, which is the one presented in this paper.

\begin{figure}
\resizebox{!}{88mm}{\includegraphics{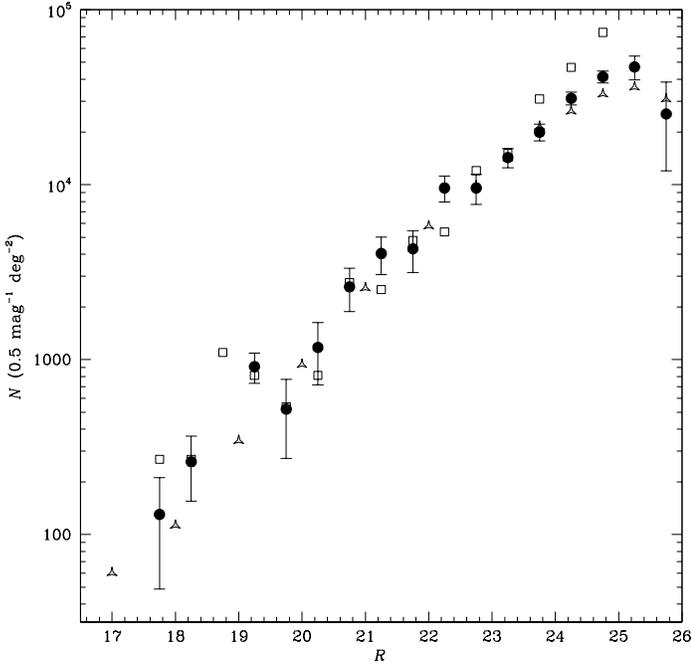}}
	\caption[]{Differential number of galaxies as a function of 
    isophotal R-magnitude for the background field (filled circles). The background counts of DPDMD (open squares) and BNTUW (open triangles) are also shown.}
         \label{FigCount_Bkgcorr}
\end{figure}

In Fig.~{\ref{FigCount_Bkgcorr}} we show the corrected R\,-\,band background
counts, together with those obtained by the Hitchhiker team at WHT (Driver et al. 1994a) and the counts from BNTUW. As seen in the figure, the corrected 
background counts agree well with each other down to $R \sim 24$, within the 
limited statistics. While we have a total background area of $27.65 
\Box\arcmin$, Driver et al. had a total area of $15.9 \Box\arcmin$. Their 
exposures with the WHT were, however, deeper in these fields. We also note 
that our counts are within the variations of other recent investigations 
such as those by Arnouts el al. (1999) and Fontana et al. (1999).

\section{The cluster luminosity function}

\subsection{Error estimation of cluster counts}

Because of both statistical and systematic errors a careful error analysis
has to be made. The standard method has been to assume 
Poisson statistics for the galaxies in the background field and cluster field, 
respectively, in some cases supplemented by an error for the
field\,-\,to\,-\,field variation caused by large scale structures. Because we have only one background
field, we cannot 
determine the field-to-field variation from our material. We therefore use the 
field\,-\,to\,-\,field statistics of BNTUW, with the characteristics of our data, 
to estimate the 
variation in the background counts as a function of magnitude (see also Trentham 1997). This value 
was added in quadrature to the Poisson variation in the cluster counts and 
the error due to the magnitude correction. The latter was estimated as 
the $1\sigma$ dispersion in a number of simulations (see 
Section~{\ref{sec:magcorr}}). The two error sources are generally of 
comparable magnitude, but the field-to-field variation is systematically 
larger for fainter magnitudes ($R > 22.5$).

\begin{figure}
\resizebox{!}{88mm}{\includegraphics{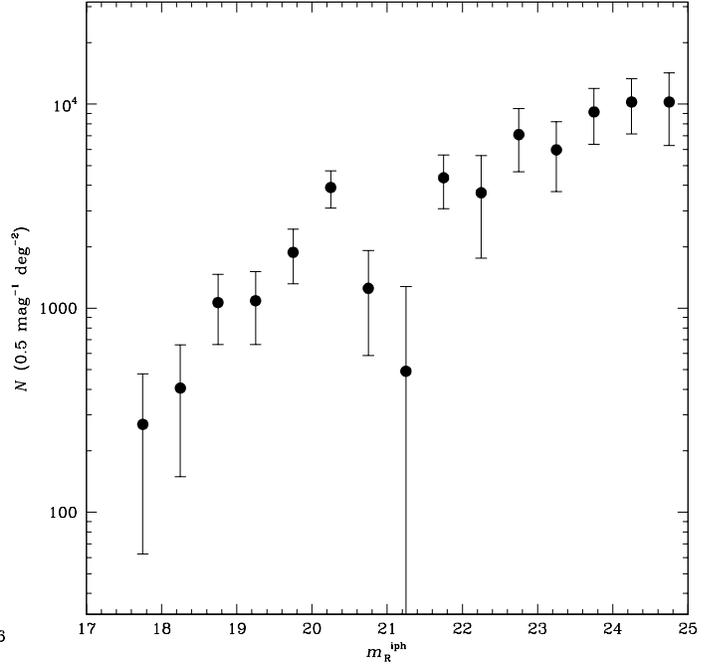}}
	\caption[]{Background-subtracted differential
              number of galaxies per 0.5 mag and square degree 
              as a function of isophotal R-magnitude for MS2255.7+2039. No 
corrections have been applied.
              }
         \label{FigCount_Cl_bkg}
\end{figure}

\begin{figure}
\resizebox{!}{88mm}{\includegraphics{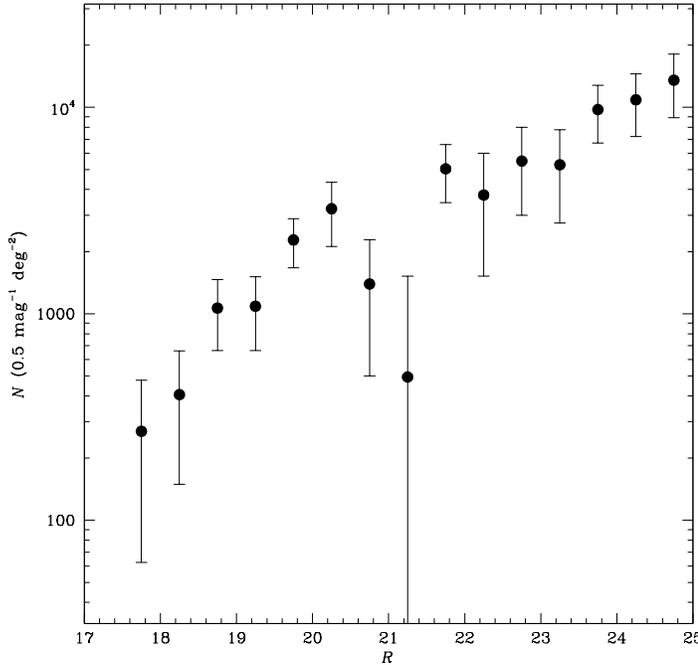}}
	\caption[]{Background-subtracted and corrected differential number 
         of galaxies per 0.5 mag and square degree as function of isophotal R-magnitude for MS2255.7+2039. 
              }
         \label{fig:FigCount_Cl_bkgcorr}
\end{figure}

\subsection{The raw LF}
\label{sec:rawLF}

By subtracting the background counts from the cluster counts, we obtain
the LF in terms of the apparent, isophotal magnitude. This 'raw' LF is 
shown in Fig.~{\ref{FigCount_Cl_bkg}}. Then we applied the
same distance modulus, including $K$-correction, as for the corrected
distribution (see below). 
The resulting distribution can be fitted by a single Schechter function with $\alpha 
= -1.4$ and $M_{\mbox{\tiny{\rm R}}}^{*} = -24$.
It is important to note that these numbers are arrived at partly because of 
the two 'low' bins at $M_{\mbox{\tiny{\rm R}}}=-20.75$ and 
$M_{\mbox{\tiny{\rm R}}}=-20.25$. If we exclude 
these points, we obtain $\alpha = -1.3$ and $M_{\mbox{\tiny{\rm R}}}^{*} 
= -22.8$. Although the shape of the LF suggests that 
the bright end would be better fitted by a Gaussian, these low values can  
be explained as a statistical effect. Both points are fitted within 
the $2\sigma$ level by a Schechter function with parameters 
$M_{\mbox{\tiny{\rm R,giants}}}^{*} = -22.5$, $\alpha = -1.0$ (see 
Fig.~{\ref{fig:ThreeLFs}}). Monte Carlo simulations of this Schechter function 
with the same total number of galaxies as observed, show that a dip similar to 
that in Fig.~{\ref{FigCount_Cl_bkg}} appears in $(10-20)\%$ of the simulations. 
We therefore conclude that this dip is of marginal significance.

The Schechter parameters can be compared to those of Valotto 
et al. (1997), who found $\alpha = -1.4$ for rich clusters and $\alpha = 
-1.2$ for poor. A linear fit for the range $-19 < M_{\mbox{\tiny{\rm R}}} < 
-17$ also yielded $\alpha = -1.3$, which is the least model dependent 
estimate of the uncorrected faint-end slope. A comparison of 'raw' LFs at 
different redshifts is, however, misleading.

\subsection{The corrected LF} \label{seccorrLF}

For an unbiased comparison with samples at other redshifts, the isophotal 
magnitudes have to be related to the total magnitude, and then translated 
into absolute magnitudes. 
For the latter we included an R-band $K$-correction given by 
Metcalfe et al. (1991), who found a $K$-correction of 0.41 magnitudes for 
E/S0 galaxies at $z = 0.288$, while the average for spirals was 0.09. 
We used a straight mean value $K = 0.15$, but realize that the true range 
may be $-0.1 \la K \la 0.41$. Because of the type dependence of the 
$K$-correction the intrinsic cluster LF is expected to be slightly 
redistributed, compared to the one observed. This redistribution should, 
however, be within the limits just mentioned. Thus, the distance modulus 
became $(m-M) = 41.47 - K$.

In order to estimate the influence of the $K$-correction on the measured slope 
we made two tests. In both cases we kept the correction of the bright 
population fixed at $K = 0.15$, but applied values of $K = 0.1$ and $K = 0.4$, 
respectively, to the dwarfs. This was done by simply assuming that all bins in 
the LF at $R > 23$ represent the dwarf population, since 
this is the region where the dwarfs dominate in the corrected LF (see 
Fig.~{\ref{fig:ThreeLFs}}). In neither case is the faint-end slope for the 
the fitting method described below affected.

The uncorrected LF presented in Section~{\ref{sec:rawLF}} represents a strict lower 
limit to the true LF. By applying the magnitude and crowding corrections, 
described in previous sections, to the cluster field and background field, 
respectively, we arrive at the LF shown in 
Fig.~{\ref{fig:FigCount_Cl_bkgcorr}}. 

While providing a useful description of the data, a Schechter fit is not 
unique, and we have tested different parametrizations of the corrected 
data. A single Schechter function with $M_{\mbox{\tiny{\rm R}}}^{*}=-23.9$ 
and $\alpha \simeq -1.43$ gives a decent fit, although there is an 
indication of a break at $M_{\mbox{\tiny{\rm R}}} \simeq -20$.

Binggeli et al. (1988) discussed the separation of the total LF 
into components for each Hubble type for the Virgo cluster. This is not 
possible for our data, because of the distance to the cluster, the lack of 
colour information, etc. We can, however, use the {\em a priori} information 
that the transition region between giants and dwarfs is at $-18 \la 
M_{\mbox{\tiny{\rm B}}} \la -16$. For comparison reasons, we add two separate 
Schechter functions, one representing giant galaxies and the other representing dwarfs. 
If the slope of the giants is fixed at $\alpha = -1.0$, 
the procedure yields $M_{\mbox{\tiny{\rm R,giant}}}^{*}=-22.8$ and 
$M_{\mbox{\tiny{\rm R,dwarf}}}^{*}=-18.9$ for the two populations. 
This is close to the corresponding values found by DPDMD ($-22.5, -19.0$). 
The main discrepancy is in the slope of the dwarf population. For MS2255
we find $\alpha \simeq -1.5$. This can be compared to A963 for which DPDMD
find $\alpha \simeq -1.8$. It is, however, 
important to note the coupling between $M^{*}$ and $\alpha$. For 
$M_{\mbox{\tiny{\rm R,dwarf}}}^{*} = -19.5$ our faint-end slope yields 
$\alpha \simeq -1.8$, while $M_{\mbox{\tiny{\rm R,dwarf}}}^{*} = -18.5$ 
results in $\alpha \simeq -1.2$ (the dwarf/giant ratio was, however, not 
constrained in these tests, as in the case of DPDMD's fit). The strong 
coupling between $\alpha$ and $M_{\mbox{\tiny{\rm R}}}^{*}$ makes it 
dangerous to draw any firm conclusions based on Schechter-function fits only. 
Instead, {\em one should directly compare any two LFs, magnitude by magnitude}, as 
shown below. In order to avoid the coupling of the Schechter-function 
parameters for the faint end of the LF, we also used a simple straight-line 
fit to the last five data bins, $-19.5 \le M_{\mbox{\tiny{\rm R}}} \le -17$ 
(as proposed by Trentham 1998b), which yielded $-1.6 \la \alpha \la -1.5$. 
This is somewhat steeper than the 'raw' LF, and is caused by the magnitude 
and obscuration corrections. This steepening of the LF is consistent with 
the bright end of the dwarf population detected in more nearby clusters 
(see discussion below). Although the procedures discussed above yield a 
formal value of the faint-end slope of $\alpha \simeq -1.5$, the uncertainties
involved in these kind of studies make us emphasize that one should not focus
on the exact value of the slope, but rather on the qualitative appearance of
the LF.

\subsection{Comparison with other clusters}
\label{comparison}

A major reason for the interest in the cluster LF is to study the 
evolutionary effects with redshift. As we have just discussed, this is 
done best by a direct comparison of the different LFs, i.e. by plotting 
them together, including all corrections. 
A problem is that the observations are in different filters and/or that the 
$K$-corrections are uncertain. A possibility to avoid some of these 
uncertainties is to compare the LFs in filters with central wavelengths 
adjusted to the redshift. In our case we note that at $z=0.288$, the R 
band corresponds to a wavelength between B and V at $z=0$. The other, more 
model dependent, method is to use a $K$-correction for a given $z$ and a 
given filter. This depends, however, on the population (section~\ref{seccorrLF}), as well as on 
evolution. Unfortunately, we are in most cases forced to use this alternative. 

\begin{figure}
\resizebox{\hsize}{!}{\includegraphics{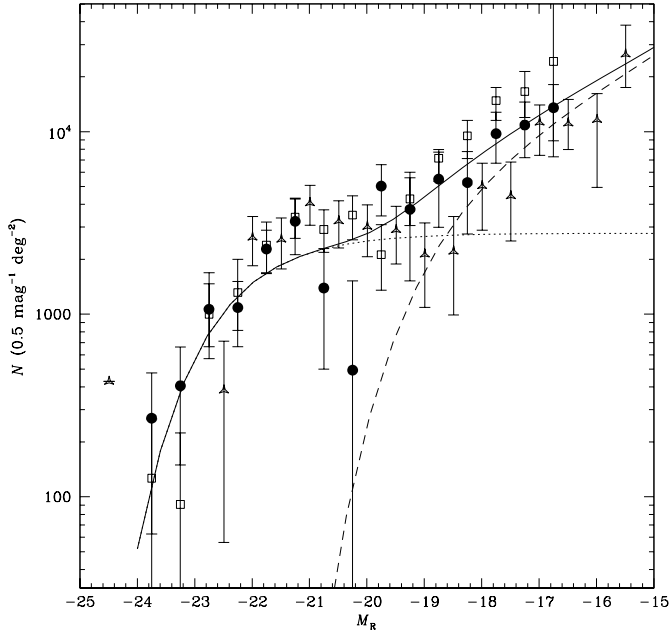}}
	\caption[]{Luminosity functions of three galaxy clusters. Filled 
circles represent MS2255.7+2039, open squares A963 (DPDMD), and open 
triangles Coma (Trentham 1998a). A963 and Coma have been normalized to 
MS2255.7+2039 for $M_{\mbox{\tiny{\rm R}}} \le -21$. The curve represents 
a combination of two Schechter functions ($M_{\mbox{\tiny{\rm R,giants}}}^{*} 
= -22.5$, $\alpha = -1.0$, dotted line; $M_{\mbox{\tiny{\rm R,dwarf}}}^{*} = -19.0$, 
$\alpha \simeq -1.5$, dashed line) and is shown here in order to guide the eye.
              }
         \label{fig:ThreeLFs}
\end{figure}

In Fig.~{\ref{fig:ThreeLFs}} we display the LF for MS2255 together with 
Trentham's (1998a) LF of Coma ($z = 0.023$) and the LF of A963 ($z = 0.206$) 
 by DPDMD, all adjusted to ${\rm H}_{\mbox{\tiny{0}}} = 50 \ {\rm km \ s}^{-1} 
{\rm Mpc}^{-1}$. Both LFs were normalized to the same level as MS2255 for 
galaxies brighter than $M_{\mbox{\tiny{{\rm R}}}} = -21$. Although a fainter 
limit for this normalization would have been preferable, this value was 
chosen to avoid influence from the low points in the LF of MS2255 at 
$M_{\mbox{\tiny{{\rm R}}}} \simeq -21$. 

It is evident that all three clusters exhibit steep slopes at the faint end, 
and there is accordingly no qualitative difference between nearby and distant 
clusters in that respect. The slope of Coma is actually as steep as that of 
A963, while MS2255 displays a somewhat flatter faint end of the LF. The steep 
slope of the Coma LF was also noted by Smith et al. (1997), who used Coma 
data from Thompson \& Gregory (1993). The steepening occurs at slightly 
different magnitudes. A963 has its break point around 
$M_{\mbox{\tiny{{\rm R}}}} = -19.5$. The steepening in MS2255 starts at 
approximately the same magnitude, while that in Coma occurs at a somewhat 
fainter magnitude ($M_{\mbox{\tiny{{\rm R}}}} \simeq -18.5$). 

The shape of the faint end of the cluster LF is a matter of controversy. 
While both nearby and more distant clusters show a steepening of the 
LF, the magnitude where this occurs differs substantially. To some extent 
this may be caused by a simple zero-point shift between filters. Most studies 
at intermediate redshifts ($0.1 \le z \le 0.2$) yield steep slopes ($-2 \la 
\alpha \la -1.7$) for $-19 \la M_{\mbox{\tiny{{\rm R}}}}$ (DPDMD; Smith et 
al. 1997; WSEC). However, the studies by Trentham (1998b, c), 
which in some cases are based on local clusters, give a more shallow slope 
($\alpha \simeq -1.4$). In some local clusters, the steep slope starts at 
$M_{\mbox{\tiny{{\rm R}}}} \simeq -14$, which is a region beyond the 
limits of present studies for clusters around $0.2 \la z \la 0.3$. One reason 
for the discrepancies can be the different correction methods applied.
It should be noted that earlier photographic investigations of 
nearby clusters also suffered from severe selections effects that work against 
low surface brightness objects. When such effects have been corrected for, the 
resulting slope is in the range $-1.5 \la \alpha \la -1.8$ (Impey et al. 
1988; Bothun et al. 1991). 

Another subject of interest is the possible existence of a universal 
cluster LF. In view of the discussion above, this is highly controversial 
and depends e.g. on the correction procedures applied. Composite LFs of 
clusters at $0.02 \simeq z \simeq 0.2$ have been presented both in B and 
in R, based on more than 15 (in B) and six clusters and groups (in R), 
respectively (Trentham 1998c, d). The composite R-band LF has a slope of 
$\alpha \simeq -1.5$ at $M_{\mbox{\tiny{{\rm R}}}} \simeq -17$, which is 
similar to what has been found here for MS2255. The steep slope of Coma 
(Trentham 1998a) seems to have been averaged out by the weighting procedure 
in the composite LF. Trentham noted, 
however, that his composite function may be valid only for the centres of 
rich clusters, i.e. regions dominated by ellipticals. 

WSEC studied two clusters, A665 and A1689, both at $z = 0.18$, 
in V and I. This study is especially interesting because it gives some 
information about the colour dependence of the LF. WSEC found rising 
LFs with breaks around $M_{\mbox{\tiny{\rm V}}} =
-19$ and $M_{\mbox{\tiny{\rm I}}} = -21$. Their V-band slope, 
after correction for incompleteness and obscuration, is very steep 
($\alpha \sim -2$), while that in the I band is significantly flatter 
($\alpha \sim -1.1$). This difference is interesting, since it indicates  
that the faintest detected galaxies indeed are blue. If this effect is real, 
one would expect an even steeper slope in B. Somewhat surprisingly, Trentham 
(1998b) did not find a comparatively steep B-band LF in his study of A665. 
Furthermore, while displaying 
very steep slopes ($\alpha < -2$), the four Abell clusters investigated by
De Propris et al. (1995) did not show any differences between B and I in 
this respect. The question of colour dependent slopes is therefore still 
unanswered. 

Few numerical simulations of cluster LFs exist. White \& Springel (1999) 
discuss in a recent paper a combination of $N$-body and semianalytical 
modelling of the cluster population. Unfortunately, they only present the 
B-band LF, for which they find a faint end slope of $\alpha = -1.2$. Although 
a direct comparison is difficult (R at $z = 0.3$ corresponds approximately to 
V at $z = 0$), this is considerably flatter than the observed slope presented 
here.

\section{Summary and conclusions}

We have observed the galaxy cluster MS2255.7+2039 ($z = 0.288$) and a
background field at similar galactic latitude with the aim of
determining the cluster LF. The isophotal magnitudes have been
corrected for light loss according to results obtained from
simulations. We have also compensated for obscuration due to bright,
apparently large, objects in the images. The resulting cluster LF 
has a fairly steep faint-end slope ($\alpha \simeq -1.5$) faintward of the 
break in the profile around $M_{\mbox{\tiny{\rm R}}} = -19$. This slope is
more shallow than some LFs found in clusters both locally and at $z \simeq 
0.2$, but similar to the slope of the composite LF derived by Trentham 
(1998c). Without focusing too strongly on the precise value of the slope, we conclude that MS2255.7+2039 exhibits a steepening LF at faint magnitudes.

The evidence for steep faint-end slopes of cluster LFs is accumulating. There 
are now a number of fairly deep CCD investigations of nearby, as well as a few 
medium distant ($z \la 0.3$) clusters, that all point to rising LFs at faint 
magnitudes. It therefore seems clear that a flat LF (i.e., $\alpha = -1$) can 
be ruled out even at intermediate magnitudes ($ -20 \la M_{\mbox{\tiny{\rm 
R}}} \la -17$). However, several questions remain unanswered. The 
uncertainties in the measured slopes are probably considerable, since 
different correction methods seem to yield deviating results, which probably 
explains the discrepancies between the LFs found for the same clusters as 
determined by different investigators (see section~\ref{comparison}). 

Because of these uncertainties, it is too early to discuss any variation of 
the faint-end slope with $z$. The accuracy of the present study only allows 
us to claim that the cluster LF is non-flat at faint magnitudes ($-19 \la 
M_{\mbox{\tiny{\rm R}}}$). The exact values of the slope and the magnitude 
where the steepening sets in are uncertain, and any trend with $z$ that may 
be present is dominated by these uncertainties. In addition, environmental differences, 
like richness or density, between clusters at the same $z$ could affect 
individual LFs, making distinctions in $z$ even more difficult to isolate. The 
uncertainties in the background subtraction is also a source of error, 
although the simulations by Driver et al. (1998) show that the faint-end 
slope of the LF can be reliably determined out to $z \simeq 0.3$ with seeing 
and depth similar to those of the present data. Nevertheless, the errors 
due to the statistical background subtraction can probably be substantially 
reduced by using photometric redshifts. Work based on this approach is in 
progress.

There are in the context of cluster LFs several important questions to 
answer in the future: Is there a universal LF for galaxy clusters at low 
redshifts, or is the steepness of the dwarf population different between 
clusters? Is there a colour dependence of the steepness of the dwarf 
population, as may be indicated in the study by WSEC? 
The K-band observations of five clusters by Trentham \& Mobasher (1998) are 
especially interesting in this context. These data were, however, not deep 
enough to
draw any conclusions about the faint end of the luminosity function. From  
the clear signs of evolution between $z = 0.5$ and the present epoch for the 
field (e.g., Lilly et al. 1995), one would expect a corresponding 
evolution in clusters. From the galaxy harassment scenario for the 
Butcher-Oemler effect (Moore et al. 1996) one may expect a larger fraction 
of low luminosity galaxies in the past, and therefore a steeper LF. 
We hope to address some of these issues in the future.

\begin{acknowledgements}

We are grateful to Helmuth Kristen for obtaining a few images of the 
cluster in September 1995 and to Steven J\"ors\"ater for some initial 
observations in 1994. We also thank Leif Festin for supplying some 
ALFOSC images that we could use to test the image quality prior to our 
observations. We are grateful to Margrethe Wold for discussions about 
completeness 
and to Tomas Dahl\'en for discussions and assistance with an additional 
consistency check. We also thank the referee, C. Gronwall, who provided 
several important suggestions that improved the presentation of this
work. Last but not least, M.~N. is very grateful to Stefan Larsson for 
discussions about statistics and related topics. The data presented here 
have been taken using ALFOSC, which is owned by the Instituto de 
Astrofisica 
de Andalucia (IAA) and operated at the Nordic Optical Telescope under 
agreement
 between IAA and the NBIfA of the Astronomical Observatory of 
Copenhagen. This 
research was supported by the Swedish Natural Sciences Research Council,  
and the G\"oran Gustafsson Foundation
 for Research in Natural Sciences and Medicine.
\end{acknowledgements}

{}
 \vfill
\eject
\end{document}